# Balancing broomsticks to a t!

David Marasco (Foothill College)

A classic question from rotational mechanics asks “Is it easier to balance a broomstick heavy-side up or heavy-side down?”[1,2] Measuring the time a student can balance an asymmetrically-weighted meterstick in two orientations is straight-forward exercise that can be used to introduce or review the t’ framework,[3] which guides lab students in comparisons of results and iterative improvement of their measurements. As a two-year college physics program, many of our students attend from other institutions, so while t’ is taught in the introductory mechanics labs, there is no guarantee that a student in a following lab course will be familiar with the material. To address this, the t’ concepts are reviewed in the first session of subsequent lab courses. The broomstick experiment is a quick and fun implementation. A spreadsheet with sample data is provided as an example.[4]

The t’ paradigm forms the core of our physics lab experience. Results of measurements are compared to either theoretical predictions or other sets of measurements, and the distance between the values is divided by their combined uncertainty. This value, labelled t’, gives the students information about their measurements and helps them determine what their next steps should be in iterating future experiments. Ideally the students cycle through several upgrades of experiments.  If the t’ value is less than one, then the two measurements are indistinguishable given their uncertainties.  If t’ is greater than 3, then one is clearly larger than the other.  Values between1 and 3 indicate inconclusive data.

As top-heavy broomsticks should be easier to balance, it is expected that after several rounds of improving experiments, t’ should be greater than 3 when comparing balancing times in the two orientations. While our mechanics labs teach this approach over the span of two three-hour sessions using the More Perfect Pendulum experiment,[3] the broomstick exercise can be done in roughly 90 minutes, ideal for a first-day lab.

After a review of the t’ system, students are instructed to tape 200g of mass at the 67cm point on wooden metersticks, which act as stand-ins for broomsticks. Secretly, one group is advised to place their mass at 80cm instead. The instructor demonstrates balancing the meterstick on their hand, and students measure how long they can balance the meterstick (*fig.1*), and to use t’ to determine if the balance times for the two orientations are distinguishable. This requires students to calculate means and standard errors for their data. A shared spreadsheet is provided for students to record their raw data. This allows the instructor to check math and to organize the data for subsequent discussion. As students generally move around to keep the meter stick balanced, groups can go outside of the lab room to either outdoors or a hallway for safety.

As the t' framework features a cycle of improvement, once the students have found their t' values they critique their experiments and look for refinements. Rather than performing an additional round of upgraded experiments, the class is asked to share their thinking on issues in their original implementation, and possible improvements. Common issues are detailed below.

Humans become better at tasks with practice,[5] so due to training effects later attempts may result in longer balance times. Alternating the two orientations between measurements would help mitigate this effect.

The exercise is a fun activity, so groups often have multiple members act as stick-balancers. Using only one person is a good control variable, as balancing skill varies (a quick look at group-to-group results will show a wide range of talent). Using multiple people is a teachable moment if it is clearly indicated in a group's data. If the data are summed across all of the group members, then the standard error will be larger, as it accounts for variation due to both luck and the possibly different skill levels of the participants. As the standard error is in the denominator of the t' ratio, this artificially depresses the t' value when compared to the t' calculated for each individual in the combined measurement.

The asymmetry of the weighted stick could be increased, either by adding more mass or moving the mass farther from the midpoint. When students suggest this, the 80cm group is revealed, and their t' result is shared out. It is typically above 3, indicating that the balancing times in the top-heavy orientation clearly longer than those for bottom-heavy.

A common student response to refining an experiment isn't a strong improvement, the suggestion is to take more data, as more trials should reduce the standard error. The instructor should point out that the standard error shrinks as the square root of the number of trials, so diminishing marginal returns sets in quickly.

The balancing times in this lab can have wide variability, which leads to the question of outliers and rejection of data. This provides for the opportunity to discuss Chauvenet's criterion,[6] which is not often a first-day topic. With the use of tables or a z-score calculator[7] students can determine the probability of the result, and consider rejecting the outlier if the product of the probability and the number of trials is less than ½. For transparency, students are told to include results both with and without the outlier.

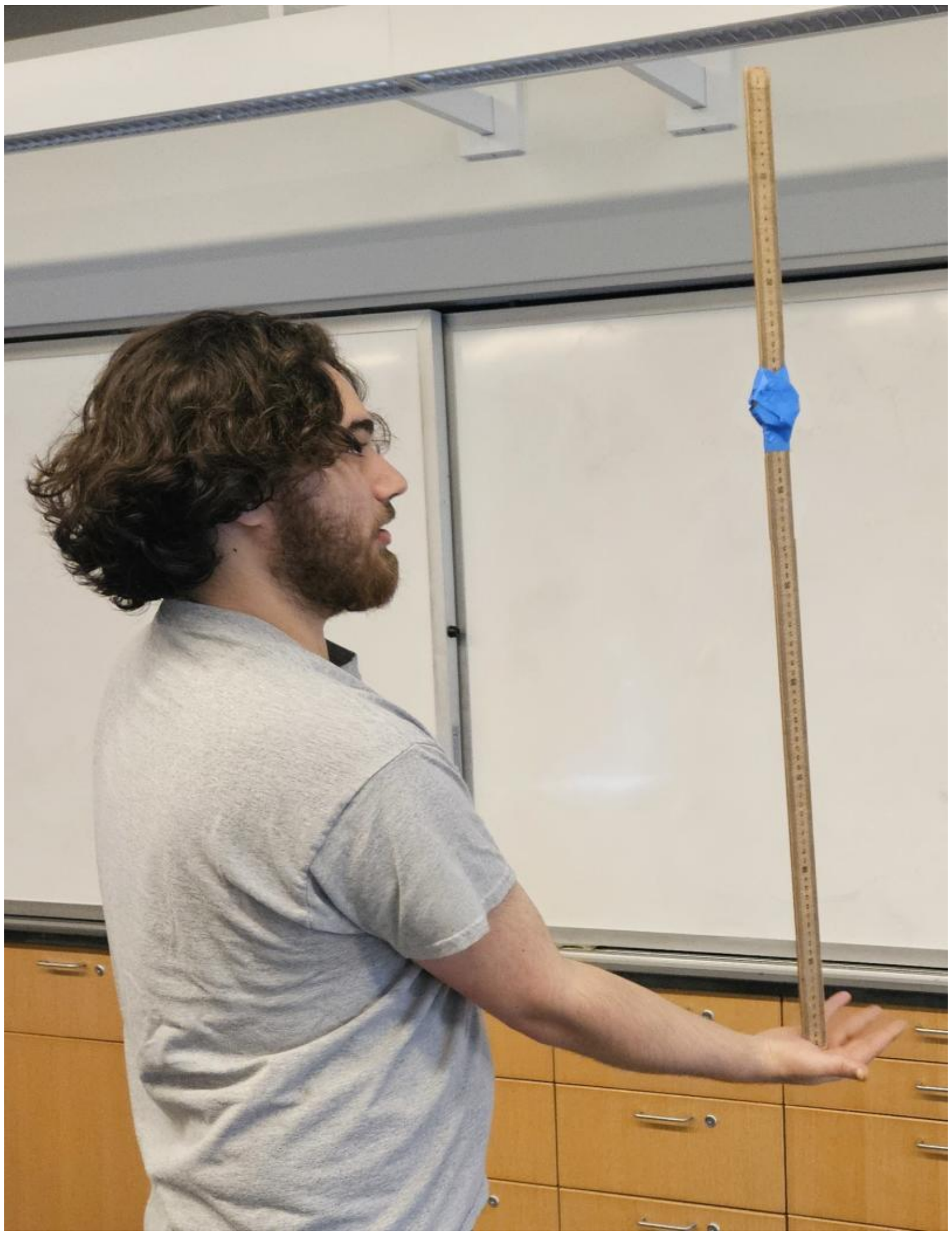

*Figure 1: A student balancing the “broomstick” in the top-heavy orientation.*